# Mechanisms of Ultrafast Charge Separation in a PTB7/Monolayer MoS$_2$ van der Waals Heterojunction

Chengmei Zhong[1,2] Vinod K. Sangwan[2], Chen Wang[1], Hadallia Bergeron[2], Mark C. Hersam[1,2,3]* and Emily A. Weiss[1,2]*

1. *Department of Chemistry, Northwestern University, Evanston, IL 60208-3113*

2. *Department of Materials Science and Engineering, Northwestern University, Evanston, IL 60208-3113*

3. *Department of Electrical Engineering and Computer Science, Northwestern University, Evanston, IL 60208-3113*

*corresponding authors. Email: e-weiss@northwestern.edu, m-hersam@northwestern.edu

**Abstract:** Mixed-dimensional van der Waals heterojunctions comprising polymer and two-dimensional (2D) semiconductors have many characteristics of an ideal charge separation interface for optoelectronic and photonic applications. However, the photoelectron dynamics at polymer-2D semiconductor heterojunction interfaces are currently not sufficiently understood to guide the optimization of devices for these applications. This manuscript is a report of a systematic exploration of the time-dependent photophysical processes that occur upon photoexcitation of a type-II heterojunction between the polymer PTB7 and monolayer MoS$_2$. In particular, photoinduced electron transfer from PTB7 to electronically hot states of MoS$_2$ occurs in less than 250 fs. This process is followed by a slower (1-5 ps) exciton diffusion-limited electron transfer from PTB7 to MoS$_2$ with a yield of 58%, and a sub-3-ps photoinduced hole transfer from MoS$_2$ to PTB7. The equilibrium between excitons and polaron pairs in PTB7 determines the charge separation yield, whereas the 3-4 ns lifetime of photogenerated carriers is limited by MoS$_2$ defects. Overall, this work elucidates the mechanisms of ultrafast charge carrier dynamics at PTB7-MoS$_2$ interfaces, which will inform ongoing efforts to exploit polymer-2D semiconductor heterojunctions for photovoltaic and photodetector applications.

**TOC Graphic:**

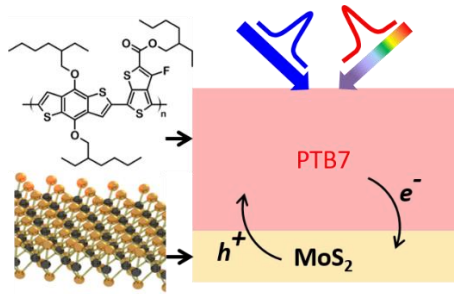

**INTRODUCTION**

The outstanding electronic and optical properties of 2D transitional metal dichalcogenides (TMDCs) have enabled demonstrations of novel field-effect transistors, solar cells, and related optoelectronic devices.[1-8] In particular, mixed-dimensional p-n van der Waals (vdW) heterostructures[9-10] comprising monolayer TMDCs and 0D/1D organic molecules or conjugated polymers have resulted in superlative thickness-normalized photovoltaic figures of merit.[11-15] Within the class of polymer-2D semiconductor heterostructures, Shastry *et al.*[16] first demonstrated bilayer heterojunction solar cells consisting of a conjugated polymer PTB7 (Poly({4,8-bis[(2-ethylhexyl)oxy]benzo[1,2-b:4,5-b']dithiophene-2,6-diyl}{3-fluoro-2-[(2-ethylhexyl)carbonyl] thieno[3,4-b]thiophenediyl}))[17-20] and monolayer $MoS_2$ that possess record-setting short-circuit current densities per unit thickness.[4, 17, 21] The internal quantum efficiency and fill factor of these devices are however only 40% and 24%, respectively, which implies that the exciton and charge carrier dynamics are far from optimal and thus require further study.

Here, we report a quantitative exploration of the ultrafast dynamics of exciton migration, charge separation, and charge recombination in PTB7-$MoS_2$ heterojunctions. The generation of photocarriers occurs on two distinct timescales: <250 fs, due to electron transfer from photoexcited PTB7 to electronically "hot" (i.e., above-bandgap) states in the conduction band of $MoS_2$; and 1-5 ps, due to both hole transfer from photoexcited $MoS_2$ to PTB7 and exciton diffusion-limited electron transfer from PTB7 to $MoS_2$. Although fast photocarrier generation mechanisms exist in PTB7-$MoS_2$ heterojunctions, our results indicate that the previously reported low internal quantum efficiency[16] is primarily due to three factors: (i) fast recombination of PTB7 polaron pairs,[22] the most reactive precursor to the charge-separated state, back to PTB7 excitons; (ii) inefficient quenching of PTB7 excitons at the PTB7-$MoS_2$ interface, resulting in prolonged PTB7 exciton lifetimes and increased probability of non-radiative recombination at MoS$_2$ surface defects; and (iii) recombination of photocarriers at defect sites in MoS$_2$.

**RESULTS AND DISCUSSION**

**Preparation and Properties of PTB7/MoS$_2$ van der Waals Heterojunctions. Figure 1A** shows the chemical structure of one unit of the electron donor polymer PTB7, and diagrams of two major photoexcited species that exist in PTB7: PTB7 excitons (PTB7*) and PTB7 polaron pairs (PTB7$^+$/PTB7$^-$). Interfacial electron transfer or hole transfer forms free polarons (PTB7$^+$/MoS$_2^-$) (not shown). To form the heterojunctions, a 10 ± 1 nm thick PTB7 layer is spin-coated onto chemical vapor deposited (CVD) monolayer MoS$_2$ after it was transferred to an ITO/glass substrate from the Si/SiO$_2$ growth substrate *via* a polymer-assisted transfer process.[16] Heterojunctions using transferred MoS$_2$ (as opposed to heterojunctions on MoS$_2$ as-grown on the original Si/SiO$_2$ substrate) are the primary focus of this study because the transparent substrate simplifies the spectroscopic analysis and is analogous to the conditions in fully fabricated solar cells. In preparing the PTB7/MoS$_2$ heterojunction thin film sample, we followed strictly the procedure developed in a previous report[16] of the device characteristics of this heterojunction. As shown in that publication, with this method, 99% of the area of the ITO substrate is covered by MoS$_2$ flakes, of which 77% are monolayer crystals, and 21% are multilayer crystals.

**Figure 1B** shows the ground state absorption spectra of two control samples, namely monolayer MoS$_2$ on ITO/glass and a 10 nm thick film of PTB7 spin-coated onto ITO/glass, and of the heterojunction film prepared as described above. The two arrows indicate the two excitation wavelengths used for the transient absorption measurements described below. **Figure 1C** shows the estimated energy level alignment at the interface between PTB7 and monolayer MoS$_2$. According to this diagram, the heterojunction is type-II, such that photoinduced electron transfer from PTB7 to MoS$_2$ and photoinduced hole transfer from MoS$_2$ to PTB7 are both energetically allowed. Also shown (where a darker shade of blue is a higher density of states) are the above-band-edge conduction band levels of MoS$_2$ that could accept photoelectrons from PTB7.

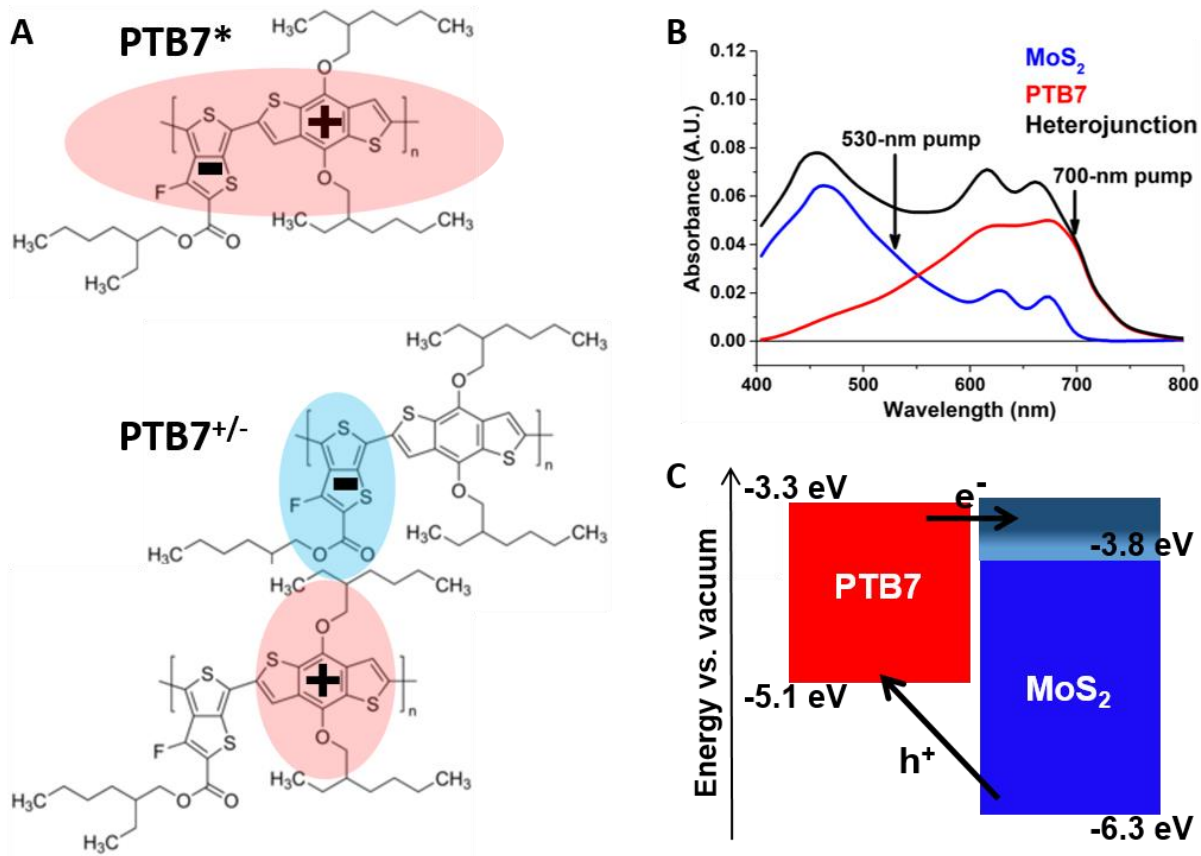

**Figure 1. A)** Chemical structure of PTB7 and charge distributions with PTB7 excitons (PTB7*) and polaron pairs (PTB7$^{+/-}$). **B)** Individual ground state absorption spectra of transferred monolayer MoS$_2$ and spin-coated PTB7 on ITO/glass, and of the heterojunction thin-film. The spectrum of the heterojunction is a linear combination of the spectra of its components. The pump wavelengths used in the transient absorption experiments are indicated by the arrows. **C)** Energies of the HOMO and LUMO of PTB7, extracted from published cyclic voltammetry measurements;[18] the valence and conduction band-edges of monolayer MoS$_2$, determined by published calculations;[21] and the above-band-edge states in the conduction band of MoS$_2$, determined by published calculations.[23, 24] Possible charge separation processes in this type-II heterojunction are electron transfer from photoexcited PTB7 to the MoS$_2$ conduction band, and hole transfer to the PTB7 HOMO from photoexcited MoS$_2$.

**Ultrafast Charge Generation in Heterojunction Thin-Films of PTB7 and Monolayer-MoS$_2$.** **Figure 2A** shows the visible regions of the transient absorption (TA) spectra of a PTB7/MoS$_2$ heterojunction film 250 fs after excitation at 530 nm, where ~40% of absorbed photons excite PTB7, and ~60% of absorbed photons excite MoS$_2$, and 700 nm, where >99% of absorbed photons excite PTB7. The main features of these spectra are the ground state bleaches (GSBs) of MoS$_2$, also present in the TA spectra of bare MoS$_2$ on ITO/glass (blue line). We make three observations from **Figure 2A**.

First, the TA spectrum of the heterojunction at 250 fs (the shortest observable time delay in our experiment) is dominated by MoS$_2$ GSB features when the sample is pumped at either 530 nm or 700 nm, even though less than 1% of the 700-nm photons are absorbed by MoS$_2$. The features of the PTB7 GSB are absent at 250 fs, but recover at later times, specifically after approximately 4 ps. The GSB of PTB7 must therefore be obscured by a positive feature for the first few picoseconds after photoexcitation at 700 nm. Comparison of the black and blue spectra in **Figure 2A** also shows that a positive feature is present in the TA spectrum of the heterojunction that is not present in the TA spectrum of isolated MoS$_2$. PTB7 has no photoinduced absorptions in this region, so we tentatively conclude that the GSB of PTB7 in the 250-fs spectrum of the heterojunction is canceled by an excited state absorption (ESA) of MoS$_2$. **Figure 2B** supports this conclusion. It shows that if we add the TA spectrum of PTB7 after pumping at 700 nm back into the TA spectrum of the heterojunction at 250 fs after pumping at 700 nm, we reproduce (approximately) the shape of the TA spectrum of the heterojunction at 5 ps, after the ESA has decayed (spectra at additional delay times are shown in Supporting Figure S1). The shape of the 5-ps spectrum is not reproduced exactly by this procedure, as the shape of the ESA of MoS$_2$ and the shape of the GSB of PTB7 are not exact mirror images.

We estimate the lifetimes of this ESA to be <250 fs (89%) and 4.1 ± 0.7 ps (11%) ps by tracking the dynamics at 642 nm of the heterojunction spectrum after pumping at 700 nm (shown in the inset of Figure 2B); at this probe wavelength, a positive feature that is *not* present in the TA spectra of either isolated PTB7 nor isolated $MoS_2$ is apparent. Since this ESA is not detectable when bare $MoS_2$ is excited directly, even with energies above its bandgap, it must be associated with a photoinduced process that occurs within our IRF (250 fs) due to the interaction of $MoS_2$ and PTB7, rather than by a straightforward above-bandgap direct excitation of $MoS_2$. Since the ESA has lifetimes of <250 fs and 4.1 ps, it can further be concluded that the ESA is due to electronically "hot" electrons in $MoS_2$ depopulated by thermalization to the band-edge. Furthermore, the ESA is more prominent when a larger percentage of incoming photons are absorbed by PTB7 with 700-nm excitation rather than 530-nm excitation, which suggests that the state that produces the ESA results from a PTB7 excited state, rather than a $MoS_2$ excited state.

All of this evidence indicates that the short-lived ESA in the TA spectrum of the heterojunction is due to electronically "hot" electrons in the $MoS_2$ conduction band injected from photoexcited PTB7. This ESA is of similar magnitude, but opposite sign, to the GSB of PTB7 and therefore approximately cancels it out for the first few ps after photoexcitation of the heterojunction. Once the hot electrons cool to the $MoS_2$ band-edge, the TA spectrum of the heterojunction includes contributions from the PTB7 GSB (**Figure 2B**). The ultrafast generation (<250 fs) of this $MoS_2$ ESA implies that the excited states that produce the ESA are near-resonant with the PTB7 LUMO at ~0.5 eV above the $MoS_2$ conduction band-edge,[17,18] which is consistent with several reported calculations of a large density of states at 0.2 – 0.6 eV above the band-edge of monolayer $MoS_2$.[23-24] Furthermore, it is reasonable that near-resonant electron transfer from PTB7 would be favored over electron transfer with a large driving force (*i.e.*, charge transfer to the $MoS_2$ conduction band edge) given the low reorganization energy of MoS$_2$ upon charging. The absence of the ESA in the TA spectrum of bare MoS$_2$ excited above the band-edge is probably due to the fast (sub-100-fs) reported thermalization of hot electrons in monolayer MoS$_2$.[25] We believe that the longer lifetimes of the ESA in the heterojunction system is due to stabilization of hot carriers injected into MoS$_2$ by Coulombic interactions with the associated hole in PTB7 at the PTB7/MoS$_2$ interface. Such stabilization has been documented previously in organic donor/acceptor heterojunction systems.[26]

The second observation from **Figure 2A** is that, at the first time-point we record, 250 fs, the peaks of both the MoS$_2$ A and B exciton GSB features in the heterojunction spectrum[27] are at a lower energy (by 7 nm or 20 meV) than in the bare MoS$_2$ spectrum (more clearly shown in the normalized TA spectra in Supporting Figure S2A). This difference in excitonic energy between bare MoS$_2$ and MoS$_2$ in the heterojunction is not in the steady-state absorption spectrum of the heterojunction but rather occurs within the instrument response function (IRF) of the laser system (250 fs) due to an ultrafast excited state process in the heterojunction. The difference persists (and in fact, gets larger due to spectral diffusion in the bare MoS$_2$ sample, as discussed below) for the entire time window of our experiment (3 ns, **Figure 2C**), so it is not due to the contribution of the ESA discussed above. It is also not due to the contribution of the PTB7 GSB to the spectrum (SI Figure S2B). Stabilization of MoS$_2$ A and B excitons is usually associated with n-doping, the formation of depletion regions, or the formation of absorptive trions in MoS$_2$.[28-31] The IRF-limited bathochromic shift of the MoS$_2$ GSB features, also seen in pentacene/monolayer MoS$_2$ heterojunctions,[30] is therefore again suggestive of photoinduced charge transfer between PTB7 and MoS$_2$ within 250 fs. The shift could result from the presence of electrons in either an electronically hot state of MoS$_2$ or at the conduction band edge of MoS$_2$ (transferred there directly or *via* the hot state). Since this heterojunction has a staggered "type-II" configuration, conduction band electrons in MoS$_2$ could be produced by electron transfer from photoexcited PTB7 or hole transfer from photoexcited MoS$_2$. The fact that the magnitude of the ultrafast shift in the MoS$_2$ GSB is the same for the 530-nm pump (where both components are excited) and the 700-nm pump (where PTB7 is primarily excited) suggests that the major source of conduction band electrons in MoS$_2$ is photoexcited PTB7.

The third observation from **Figure 2A** is that the ratio of the amplitudes of the A and B exciton bleaches in the heterojunction spectra (A/B ratio) is different than in the bare MoS$_2$ sample, a characteristic that was also reported in pentacene/MoS$_2$ heterojunctions.[30] This difference persists for our entire measured time window (3 ns, **Figure 2C**), so it is not due to the "hot-electron" ESA. It is also not due to the contribution of the PTB7 GSB features (see Supporting Figure S2B). Similar to the energies of the A and B excitons, the A/B ratio is influenced by interactions between these excitons originating from spin-orbital splitting in the valence band of monolayer MoS$_2$.[32] We therefore suggest that photocarriers in MoS$_2$, formed from ultrafast interfacial charge transfer, decrease the number of these interactions, and thereby changes the A/B ratio.

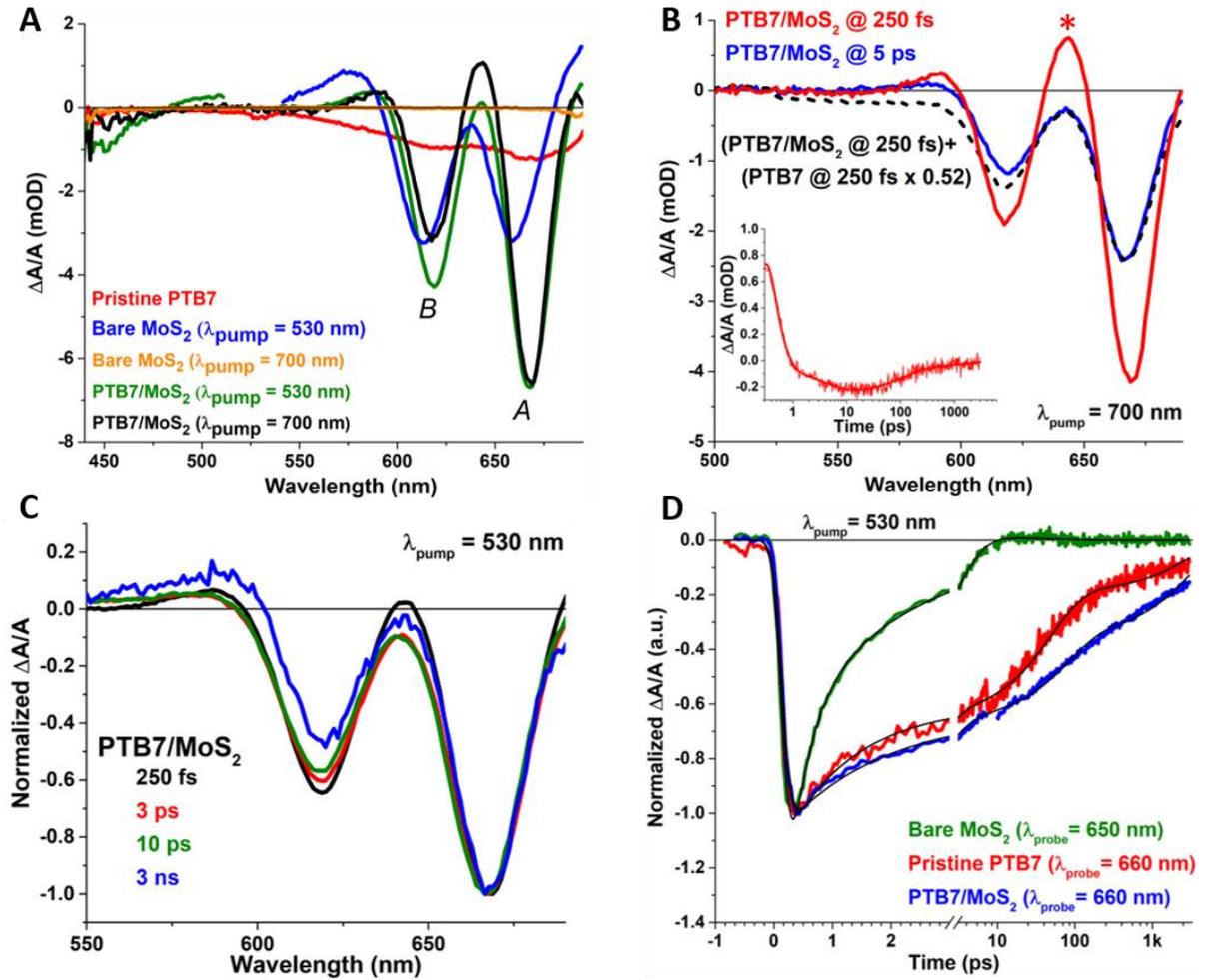

**Figure 2. A)** TA spectra of a 10-nm-thick film of bare PTB7 (pumped at 700 nm), a bare $MoS_2$ monolayer film (pumped at 530 nm and at 700 nm), and a PTB7/$MoS_2$ heterojunction film (pumped at 530 nm and 700 nm) at a time delay of 250 fs. The whited-out region between 510 nm and 540 nm in the 530-nm pump data is not accessible due to pump scatter. The $MoS_2$ A and B exciton peak positions are marked. **B)** TA spectrum of the PTB7/$MoS_2$ heterojunction film at time delays of 250 fs and 5 ps, compared to the sum of the heterojunction TA spectrum and the bare PTB7 TA spectrum at 250 fs, all excited at 700 nm. *Inset:* Kinetic trace extracted from the TA spectrum of the PTB7/$MoS_2$ heterojunction film at 642 nm (marked with an *), the peak of the ESA of hot electrons in $MoS_2$, fit with a sum of exponential functions convoluted with an instrument response function. **C)** Normalized TA spectra of a PTB7/$MoS_2$ heterojunction film in the region of the GSBs of the $MoS_2$ excitons, at time delays of 250 fs, 3 ps, 10 ps, and 3 ns after excitation at 530 nm. **D)** Normalized kinetic traces extracted from the TA spectra of bare PTB7, bare $MoS_2$, and PTB7/$MoS_2$ heterojunction films, at the wavelengths indicated in the legend, and fit with a multi-exponential function convoluted with the instrument response function (IRF). The excitation wavelength is 530 nm. All fitting parameters are listed in Supporting Table S1. All TA spectra are acquired at a pump intensity of 10 µJ cm$^{-2}$ at 530 nm or 7.5 µJ cm$^{-2}$ at 700 nm to ensure the same initial excitation density ($1\times10^{19}$ cm$^{-3}$) in all of the thin-film samples.

**Figure 2C** presents evidence that hole transfer from MoS$_2$ to PTB7 also occurs on ultrafast timescales in these heterojunctions, not necessarily in <250 fs, but certainly faster than 3 ps. Specifically, in bare monolayer MoS$_2$ (SI Figure S3), the GSB shifts by 6 nm (17 meV) to a higher energy level with a time constant of 3.0 ± 0.5 ps due to strong exciton-exciton interactions in the quantum-confined monolayer.[30, 33] This shift is absent in the spectrum of the PTB7/MoS$_2$ heterojunction, **Figure 2C** and Figure S3 of the SI. This result suggests that exciton dissociation, which must be in the form of hole transfer from MoS$_2$ to PTB7, occurs faster than the exciton-exciton interaction in MoS$_2$ (*i.e*., faster than 3 ps) in the heterojunction.

**Figure 2D** shows the kinetic traces extracted from the TA spectra of PTB7, MoS$_2$, and the PTB7/MoS$_2$ heterojunctions, all pumped at 530 nm. The kinetic traces are extracted at the peak of the A-exciton GSB of MoS$_2$ for all the samples that contain MoS$_2$. For the bare PTB7 sample, the kinetic trace is extracted at the peak of the PTB7 GSB. SI Table S1 lists time constants determined from these fits along with a sum of exponential components convolved with the IRF of the laser system, and assignments of these time constants to specific physical processes. Here, we highlight that the lifetimes of these GSBs are longer in the heterojunctions (blue and purple traces) than in any of the isolated materials. The average lifetime of the GSB of transferred MoS$_2$ in the heterojunction is a factor of five longer than in the bare MoS$_2$ and a factor of two longer than the bare PTB7. This result is consistent with photoinduced charge separation in the heterojunction, because the lifetimes of photogenerated charge carriers (PTB7$^+$ and MoS$_2^-$), which also bleach the ground state absorption, are longer than those of Coulomb-bound excitons and polaron pairs, the dominant photoexcited states in the isolated materials.

The evidence for interfacial charge transfer in the PTB7/MoS$_2$ heterojunction on the femtosecond-to-picosecond timescale outlined above is supported by species-specific signals in the near-infrared (NIR) region of the TA spectrum shown in **Figure 3A** (red line). The NIR TA spectrum of the heterojunction is dominated by the signals from PTB7, specifically PTB7 excitons (PTB7*), centered at ~1300 nm, and PTB7 polarons (PTB7$^+$), centered at ~1150 nm.[34] We employ global analysis to produce the basis spectra (**Figure 3A**, blue, black) and kinetics (**Figures 3B, C**) of PTB7* and PTB7$^+$ within the heterojunction and bare PTB7 films. PTB7$^+$ can be in the form of free carriers or polaron pairs (**Figure 1A**); these two forms are not spectroscopically distinguishable but can, in principle, be distinguished by their dynamics.

**Figure 3B** shows that, in both the bare PTB7 thin film and the PTB7/MoS$_2$ heterojunction, PTB7* and PTB7$^+$ form instantaneously (<250 fs) upon 700-nm excitation. This result is expected regardless of the degree of interfacial charge transfer since polaron pairs in PTB7 are directly generated within 100 fs.[35-36] We make three important additional observations from **Figure 3B.** First, in the kinetic trace of the heterojunction, there is a non-instrument-limited growth in the population of PTB7$^+$ in 6.5 ps ($\tau_r$ of PTB7$^+$ in Supporting Table S1) that is matched by a 5.3-ps decay in the population of PTB7* ($\tau_1$ of PTB7* in Supporting Table S1). These components are absent in the kinetics of bare PTB7. Given that 5-6 ps falls well within the range of reported values for exciton diffusion times for this polymer,[37-38] these components almost certainly correspond to dissociation of PTB7* to form PTB7$^+$/MoS$_2^-$ with a rate limited by the diffusion of PTB7* to the interface. PTB7$^+$ formed by this mechanism accounts for ~10% of the total population of PTB7$^+$ in the heterojunction when excited at 700 nm.

Second, the average lifetime of PTB7$^+$ is a factor of 1.3 longer in the PTB7/MoS$_2$ heterojunction than in the bare PTB7 film. This result is consistent with the longer lifetimes of the MoS$_2$ GSBs in the heterojunction than in the bare MoS$_2$ film due to the formation of PTB7$^+$/MoS$_2^-$ charge-separated states and suggests that some population of PTB7 polaron pairs is converted to free carriers in the heterojunction by interfacial charge transfer.

Third, upon excitation at 700 nm, the average lifetime of PTB7* is a factor of two longer in the heterojunction than in the bare PTB7 film. Specifically, there exists a decay component, $\tau_3$, in the fit of the PTB7* kinetic trace in the heterojunction that is not present in the bare PTB7 sample (see Supporting Table S1). This result is unusual for a type-II heterojunction (exciton lifetimes in individual semiconductors constituting the heterojunction are typically shortened relative to the isolated films), and suggests that the heterojunction pushes the equilibrium between polaron pairs and excitons (**Figure 1A**) toward the longer-lived polaron pair state – possibly by depleting the population of polaron pairs through interfacial charge transfer – effectively lengthening the lifetime of the exciton. This hypothesis is supported by the fact that, upon pumping the heterojunction at 700 nm, the decay of PTB7* acquires a long-timescale component ($\tau_3$) that is close to the value of $\tau_3$ for PTB7$^+$. The presence of this equilibrium in the heterojunction suggests that a large portion of polarons in PTB7 remains in the form of non-charge-carrying polaron pairs (that can recombine to excitons) rather than PTB7$^+$/MoS$_2^-$, a sign of inefficiency in the overall charge separation process. The fraction of PTB7$^+$ that is in equilibrium with PTB7* is approximately equal to the fractional amplitude of PTB7* - PTB7$^+$ coupling term ($\tau_2$ of the PTB7$^+$ kinetics of the heterojunction in Supporting Table S1), which is ~50%.

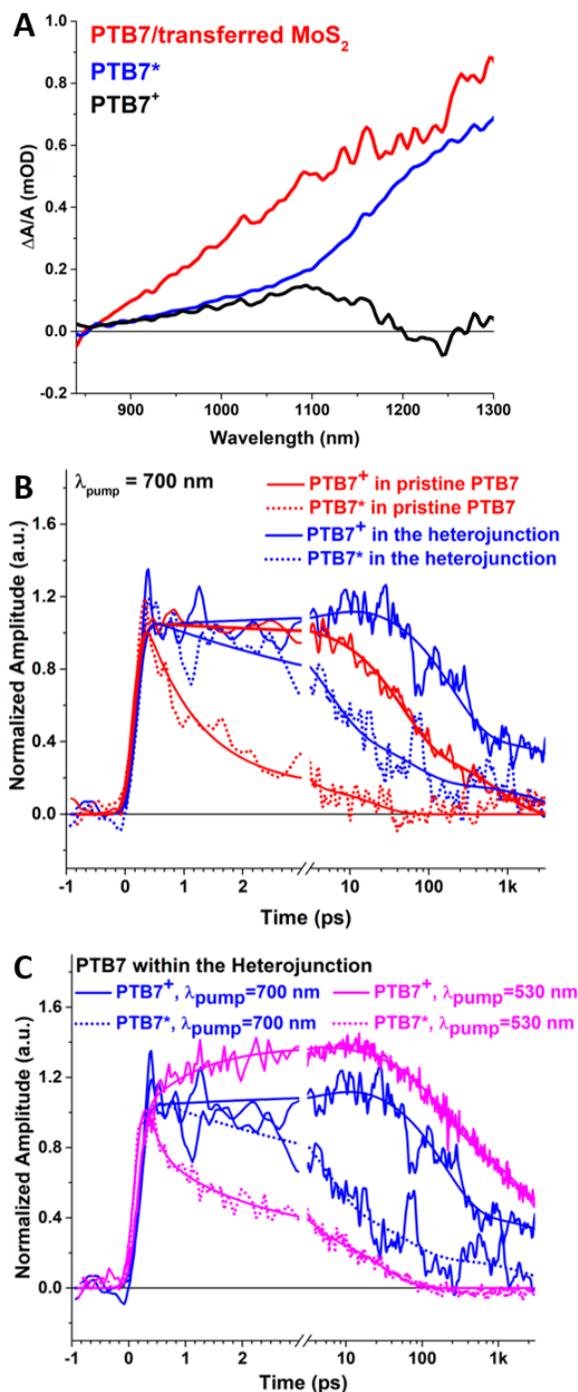

**Figure 3. A)** TA spectrum of the PTB7/MoS$_2$ heterojunction in the NIR region, 1 ps after excitation with a 700-nm pump (red), normalized to its peak value at ~1200 nm. The two major components of this spectrum (PTB7$^+$ and PTB7*) are obtained by global analysis as described in the text. **B)** Normalized kinetic traces of the PTB7$^+$ (polaron) and PTB7* (exciton) species within the PTB7/MoS$_2$ and the bare PTB7 sample after pumping at 700 nm, obtained from the deconvolved TA spectra in panel A. **C)** Comparison of the normalized kinetic traces of PTB7$^+$ and PTB7* species in the PTB7/MoS$_2$ heterojunction after pumping at 530 nm *versus* 700 nm.

**Figure 3C** compares the dynamics of the PTB7 excited states within the heterojunction when pumping at 700 nm *versus* 530 nm; the 700-nm data is repeated from **Figure 3B**. We first observe that, upon 530-nm excitation, there is again a growth in the population of PTB7$^+$ that is concurrent with a decay in the population of PTB7*. This growth component is, however, both more prominent (20% *vs*. 10% of total PTB7$^+$ population) and faster ($\tau_r$ = 1.0 ps *vs*. 6.5 ps, Supporting Table S1) than when pumping at 700 nm. We suspect that above-bandgap excitons generated in PTB7 by pumping at 530 nm diffuse faster, possibly because higher-energy PTB7* states are more spatially delocalized and dissociate more readily than band edge excitons.[39-40]

**CONCLUSIONS**

**Figure 4** summarizes the excited state and photocarrier generation and recombination processes within the PTB7/MoS$_2$ van der Waals heterojunction upon excitation of PTB7 at 700 nm and 530 nm and MoS$_2$ at 530 nm. The most striking result of our study is the occurrence of ultrafast transfer of electrons from photoexcited PTB7 to an electronically hot conduction band state of MoS$_2$, which occurs in less than 250 fs and manifests an excited state absorption (ESA) of MoS$_2$ that we do not detect upon direct excitation of MoS$_2$, even at energies above the bandgap. We propose that, unlike hot electrons in bare MoS$_2$, which have lifetimes of <100 fs, well within our IRF, hot electrons in MoS$_2$ delivered by photoinduced electron transfer from PTB7 are stabilized by the holes left behind in PTB7. The IRF-limited charge transfer (<250 fs) observed in PTB7/MoS$_2$ heterojunction is possibly of the same origin as the ultrafast charge transfer process (<100 fs) reported in polymer/fullerene heterojunctions,[37-38, 41-43] which involves delocalized excitons [41] and electrons [42] in the organic semiconductor that do not need to diffuse to the interface before transferring to MoS$_2$-localized states. There is additional generation of photocarriers in the heterojunction by exciton diffusion-limited interfacial dissociation of PTB7 excitons in 1-5 ps (depending on the initial energy of the exciton) with an apparent yield of ~58%, and interfacial dissociation of $MoS_2$ excitons, *via* hole transfer from photoexcited $MoS_2$ to PTB7, within 3 ps.

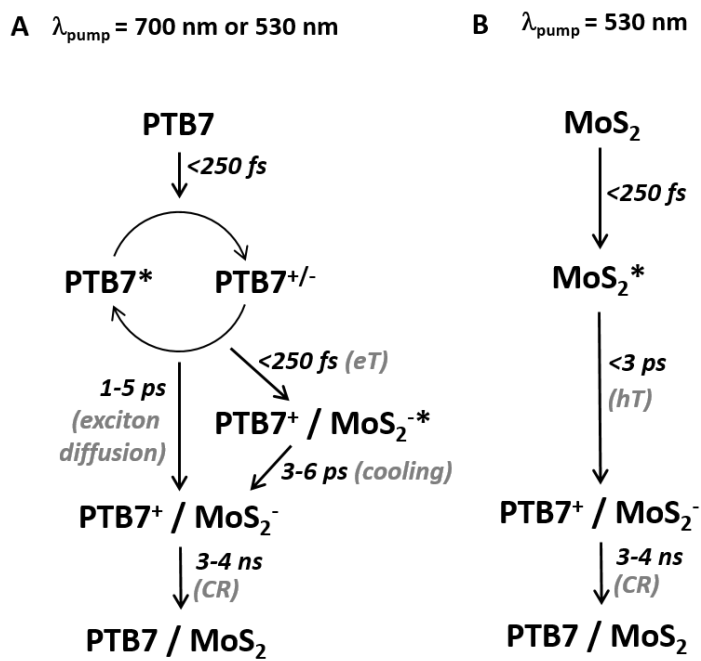

**Figure 4.** Schematic diagram of the excited state processes of each component of a PTB7/$MoS_2$ heterojunction with different excitation wavelengths ($\lambda_{pump}$), derived from fits of kinetic traces extracted from visible and NIR transient absorption spectra of the junctions presented in Figures 2 and 3. Supporting Table S1 compiles the time constants extracted from multi-exponential fits of these kinetic traces.

Previous work[16] has shown that photoexcitation of PTB7 has minimal contribution to the photocurrent in devices based on this heterojunction. We suspect this result is due to the fast equilibrium between PTB7 polaron pairs (the most probable precursors to the charge-separated state) and PTB7 excitons, and the trapping of photoelectrons from PTB7 in interfacial defect states of $MoS_2$, especially if PTB7 donates to above-band-edge states of $MoS_2$ where more trap sites are energetically accessible. Defects in $MoS_2$ will also decrease the apparent contribution of PTB7 to the photocurrent by trapping carriers even if they are injected into the lowest-energy $MoS_2$ conduction band states. It is also possible, given the energy of the first triplet excited ($T_1$) state of PTB7,[44] that back electron transfer from $MoS_2$ to PTB7 to form triplet states in PTB7 serves as an additional photocarrier recombination pathway in the heterojunction, as it does in $MoS_2$/phthalocyanine heterojunctions.[45]

Even though we find that charge injection across this van der Waals interface is extremely fast, minimization of interfacial defects will probably be necessary to inhibit recombination of photocarriers, and thereby increase the photocurrent density and fill factor of solar cell devices based on this heterojunction. Possible pathways to interfacial defect mitigation include the use of as-grown $MoS_2$ instead of transferred $MoS_2$ and chemical passivation of $MoS_2$.[46-48]

**Acknowledgements.** This work was primarily supported by the Materials Research Science and Engineering Center program of the National Science Foundation (DMR- 1720139). Chemical vapor deposition growth of $MoS_2$ was supported by the National Institute of Standards and Technology (NIST CHiMaD 70NANB14H012).

## EXPERIMENTAL METHODS

**Sample Preparation.** Large-area monolayer $MoS_2$ films were grown directly on $Si/SiO_2$ wafers with an oxide thickness of 300 nm by chemical vapor deposition (CVD) and then transferred onto ITO/glass as reported previously.[16] A 10 ± 1 nm thick (measured by atomic force microscopy) PTB7 thin-film was directly spin-coated onto the transferred $MoS_2$ on ITO/Glass (bare PTB7 chlorobenzene solution with a concentration of 3 mg/mL, spin-coated at 4000 rpm) to form a bilayer heterojunction thin-film. After fabrication, the thin-film samples were loaded into 2 mm quartz cuvettes in the $N_2$ glovebox and sealed before taking out into ambient atmosphere for TA measurements.

**Transient Absorption Spectroscopy in Visible and Near-Infrared.** Visible and NIR transient absorption spectroscopy measurements were performed following previously published protocols.[30, 49]

**Global Analysis and Spectral Deconvolution**. Global analysis was performed by first fitting the raw TA kinetic traces at each probe wavelength with two exponential decay terms. Two representative decay times were then chosen by averaging the decay terms in all probe wavelengths. The two decay times were then held constant to perform the second round of fittings on all probe wavelengths, where the amplitudes were allowed to vary, producing spectra representative of each exponential decay component.[38]

The amplitude kinetic trace deconvolution of different species from the raw TA dataset was performed by a linear combination of two chosen basis spectra using MATLAB software.[50] The output from this method is a species-associated kinetic dataset, which contains respective contributions of these two basis species to the overall raw TA spectrum at each time point.